\documentclass[twoside,11pt]{article}
\usepackage{amssymb,amsmath}

\usepackage{lineno}
\usepackage{hyperref}
\modulolinenumbers[5]

\usepackage{amssymb, amsmath}
\usepackage{graphicx, subfigure, wrapfig}
\usepackage{amsfonts}
\usepackage[english]{babel}

\def\0{{\bf 0}}
\def\a{{\bf a}}
\def\b{{\mathbf{b}}}

\def\p{{\mathbf{p}}}
\def\q{{\mathbf{q}}}

\def\s{{\mathbf{s}}}
\def\F{{\mathbf{F}}}
\def\B{\mathbf{B}}
\def\I{\mathbf{I}}
\def\M{\mathbf{M}}
\def\N{\mathbf{N}}
\def\R{\mathbb{R}}
\def\S{\mathbf{S}}
\def\U{\mathbf{U}}

\def\t{{\bf t}}

\def\u{{\bf u}}

\def\x{{\bf x}}

\def\I{\mathrm{I}}

\newtheorem{theorem}{Theorem}[section]
\newtheorem{remark}[theorem]{Remark}
\newtheorem{proposition}[theorem]{Proposition}
\newtheorem{lemma}[theorem]{Lemma}

\newtheorem{definition}[theorem]{Definition}

\begin{document}

\title{Block regularisation of the logarithm central force problem}

\author{Archishman Saha\footnote{
Department of Mathematics and Statistics, University of Ottawa,  Canada
Email: asaha018@uottawa.ca} \,\,and  
Cristina Stoica\footnote{Department of Mathematics, Wilfrid Laurier University,  Waterloo, N2L 3C5, Canada. Email: cstoica@wlu.ca}}

\maketitle

\begin{abstract}
\noindent
The logarithm  function is the gravitational potential in $\mathbb{R}^2$. We prove that the  logarithm central force problem is block regularizable, that is, the (incomplete) flow may be continuously extended over the singularity at the origin after an appropriate re-parametrization. 

\end{abstract}

%\tableofcontents

\section{Introduction}

%One of the oldest area of research in mathematics is celestial mechanics. Understanding the motion of the Sun, planets and stars inspired breakthrough achievements  in science and mathematics, from  the laws of Newton, Laplace formulae, to the modern perspective in non-linear dynamics of Poincar\'e and Arnold.

   % It is then of interest to study mathematically the logarithm central force problem and, in particular,  understand the dynamical behaviour near collision. 

During by the space race of the second half of the last century research in celestial mechanics experienced an invigorating period.  A multitude of problems had to be solved at the theoretical and practical level. Amongst these, the singularity at collision in the equations of motion lead  to extenuatory numerical approximations, as a smaller and smaller integration time interval was required to maintaining the model accuracy. \cite{StSc71}.
%
%
%: the equations of motion of the gravitational central force problem are singular at the origin. 
%Consequently, numerical approximations were difficult to complete, given that a smaller and smaller integration time interval was required for maintaining the accuracy  near the singularity \cite{StSc71}.
   In this context,  regularizing procedures became necessary, that is methods of transforming the equations, so that the ``new" flow, identical up to a parametrization to the initial one, could be extended at least continuously over the  singularity.
%Consequently,   a smaller and smaller integration time interval was required for maintaining the accuracy  near the singularity \cite{StSc71}, obstructing numerical approximations.  

%\medskip 
 %  In this context,  regularizing procedures became necessary. By regularization, we understand  transforming the equations via changes of variable and time rescalings, so that the new flow, identical up to a parametrization to the initial one, may be extended continuously over the  singularity.

\medskip
   The  elegant work of Levi-Civita, completed at the beginning of the 20th century \cite{LC13},  was brought in the  limelight in \cite{StSc71}: on a negative fixed level of energy, the planar Kepler problem  may be transformed into to a harmonic oscillator  with frequency depending on (the negative) energy. 
  % 
   % The  elegant work Levi-Civita work from at the beginning of the 20th century \cite{LC13}  was brought in the  limelight \cite{StSc71}: by an elegant change of variable and time re-pametrization, the particle motion  transforms into a harmonic oscillator  with frequency depending on (the negative) energy. 
   %The regularisation for the spatial problem was completed  in the mid sixties by Kustaanheimo and Stiefel  by embedding the problem in $\mathbb{R}^4$ and involving the algebra of quaternions \cite{KS65}. %
% \medskip
%
    The method of Levi-Civita was extended to include homogeneous potentials of the form $\displaystyle{V(r)=-1/r^{2-2/n}},$ $n\geq 2,$ $n\in \mathbb{N}$ ($r$ being the distance from the particle to the centre) by  McGehee \cite{McGe81}.  This was achieved while comparing two different regularisation methods: the ``branch regularization" of Sundman \cite{Su07} and ``block regularisation", or ``regularisation by surgery" designed by  Conley \& Easton   \cite{CoEa71}. The first  extends the double collisions as convergent power series in time in the complex plane. The second uses orbits that pass nearby the origin to extend, at least continuously, the flow past collision. 
    %For this purpose, the singularity at the origin is blown-up into an invariant \textit{collision} manifold pasted into the phase space for all levels of energy, and the re-parametrized flow is complete. 
    The regularisation of the Kepler problem and connections with various mathematical physics fields was, and continues to be, the subject of a multitude of papers: \cite{Mo70, Mil83, HedL12, GBM08, San09, LZ15, vdM21, ChHs22}, to name a few.

     \medskip

    The gravitational potential  $(-1/r)$ is  the fundamental solution of the Laplace equation in three dimensions. In two dimensions, the natural gravitational potential, understood as a solution of the same equation,   is  $V(r)=\ln r$.     
    Logarithm potentials are used in astrophysics,  in  attempts  to construct self-consistent models of galaxies; see for example  \cite{BiTr87, MESc89}, and more recently,  \cite{BBP07, VWD12}. 
    When compared to its Newtonian counterpart, the logarithm central force problem is also integrable, but all its trajectories  are bounded (there is no escape to infinity) and not necessarily closed. As an attractive law,  the logarithm law  pull   is weaker than in any law of the form $-1/r^{\alpha},$ $\alpha>0$ at close range, but stronger at large range.   Not much is known about ``logarithm" $n$-body problems, except for astrophysicists' numerical simulations  \cite{BiTr87, MESc89, BBP07, VWD12}.    In the context of  regularisation via smoothing, in \cite{CaTe11} it is proven that in the logarithm central force problem  solutions ending/emerging in/from collision may be replaced with transmission trajectories.
    A  study concerning the anisotropic two body problem is given in \cite{StFo03}.   No studies on the problem for $n\geq 3$ seem to exist.
    %Not much is known about ``logarithm" $n$-body problems, except for the numerical simulations done by astrophysicists. A study concerning the anisotropic two body problem is given in \cite{StFo03},  but no studies on the problem for $n\geq 3$ seem to exist.

  \medskip
    
 Loosely speaking, an incomplete flow is   \textit{block regularizable} if solutions that asymptotically end in the singularity set (in our case the collision set)  are in a bijective correspondence   to solutions  asymptotically leaving  the singularity set.  First 
 the singularity set is blown up  into an invariant manifold pasted into the phase space, so that the transformed flow is complete.   %Then one must prove that the map that associates orbits that enter and exit a neighbourhood of the invariant manifold replacing the singularity can be at least continuously extended to orbits that asymptotically tend  to/leave from invariant set.
   Then one defines \textit{the map across the block} that associates solutions that enter a neighbourhood of the invariant manifold replacing the singularity to solutions that exit the same neighbourhood. If this map  can be at least continuously extended to solutions that asymptotically tend  to/leave  the invariant set, then the transformed flow is called to be trivializable, and the initial incomplete flow is  block regularizable.

    \medskip
   In this paper we prove that the  logarithm central force problem is block regularizable. As in McGehee   \cite{McGe81}, the blown up  singularity set, called the \textit{collision manifold},   is an invariant torus pasted into the phase space for all levels of energy. Since  the logarithm  function  does not allow the use of the analytic transformations as in \cite{McGe81}, for the blown-up procedure we use  ${\cal C}^{1}$ transformations similar to those in  \cite{StFo03}. However, the loss of smoothness does not impede our further analysis. The main result is proven of the  bases of two facts: first, 
    %the angular momentum conservation  receives a convenient regularised expression and 
    collisions are possible only for zero angular momenta 
    and  second, on the torus collision manifold, the flow takes a very simple form. 
    %As the angular momentum $c\to 0$, trajectories that almost reach the collision manifold match those that exit and at the limit $c=0$ the flow can be extended continuously.  
    
      \medskip

The work is organised as follows: in Section \ref{sect:iso_block} we briefly introduce Conley and Easton's \cite{CoEa71} theory on trivializable isolating blocks for complete vector fields. In Section   \ref{sect:sing_block} we define block regularisation for singular of vector fields. In the next sections we discuss the logarithm central force problem, regularise the equations and the integrals of motion, and  prove our main result.

\section{Invariant manifolds, isolating blocks and trivializable flows}
\label{sect:iso_block}

 Consider a time-reversible system 
\begin{equation}
\dot \x =\F(\x)\,,\,\,\,\, \x \in \M
\label{eq:1}
\end{equation}
where $\M$ is a smooth manifold in $\mathbb{R}^n$, and $\F$ is ${\cal{C}}^\infty$ on its domain.

\medskip
We assume that the flow  of \eqref{eq:1} is complete and denote it by $\boldsymbol \Theta (\x, t)$, $\x \in \M$. A compact flow-invariant set $\N \subset \M$ is called \textbf{isolated} if there is an open set $\U \supset \N$ such that if  $\boldsymbol\Theta (\x, \mathbb{R}) \subset \U$ then $\x \in \N.$ The set $\U$ is called an \textbf{isolating neighbourhood} for $\N.$

\begin{figure}[h!]
\centering
  %     \subfigure[ Cases $C>0,$ $D>0$ for $2C-3D<0$ (top), $2C-3D=0$ (middle), $2C-3D>0$ (bottom)  ]
  %    {
       %    \includegraphics[angle=0,scale=0.3] {C>0-and-D>0.jpg}
   %        }
           \includegraphics[angle=0,scale=0.25] {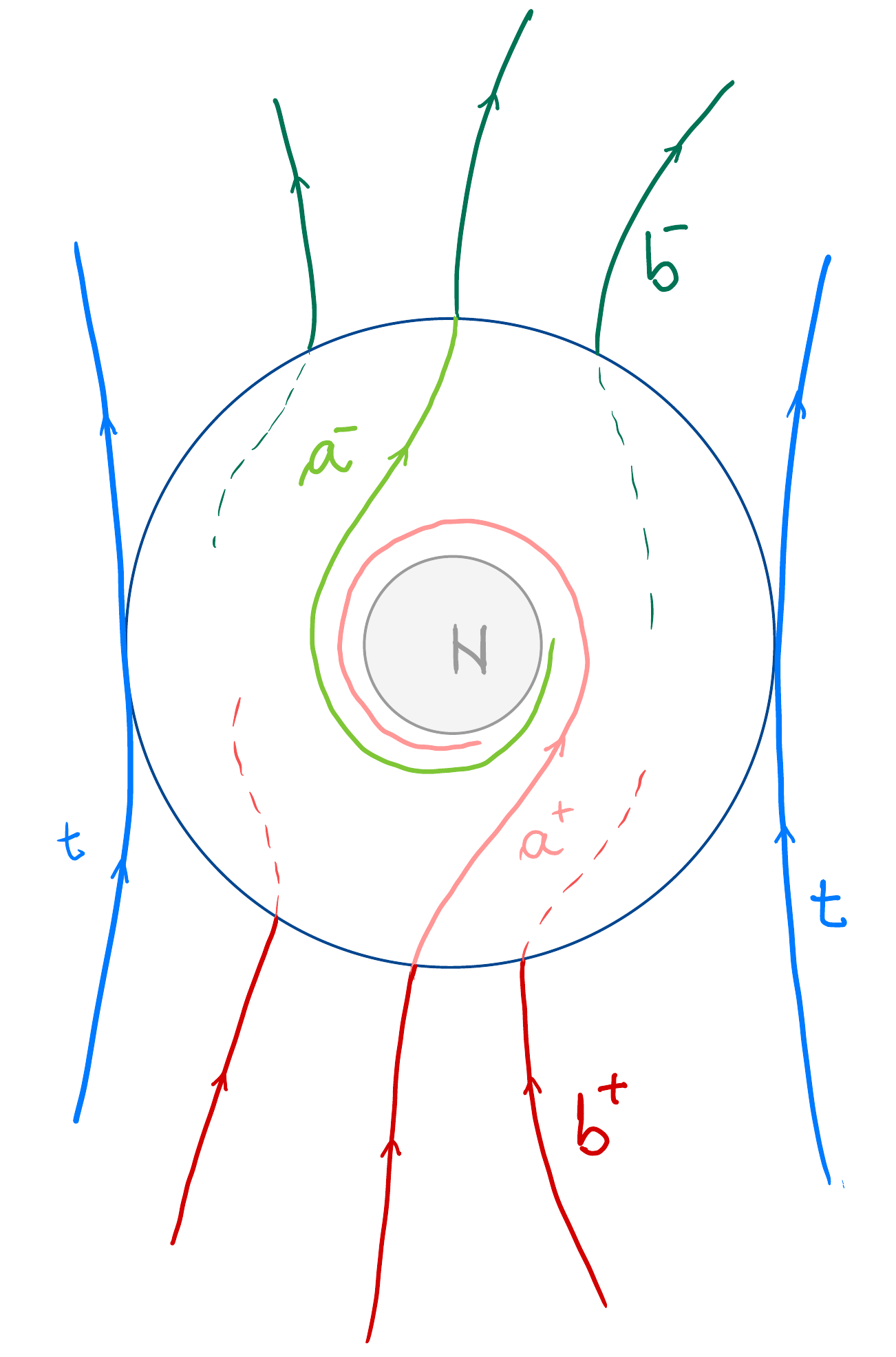}
           \caption{An isolating neighbourhood $N.$} 
         \label{isolated}
\end{figure}

\medskip
Consider $\B$ a compact subset of $\M$ with non-empty interior and assume that $\b:= \partial \B$ is a smooth submanifold of $\M$. Define (see Figure \ref{isolated})
\begin{align}
&\b^+:= \{\x \in \b\,|\, \boldsymbol \Theta (\x, (-\varepsilon\,,0)) \cap \B =\phi  \,\,\text{for some}\, \varepsilon>0\}
%= \text{initial conditions in $\b$ that in the past were not in $\b$}
\label{eq: b_plus}\\
&\b^-:= \{\x \in \b\,|\, \boldsymbol \Theta (\x, (0\,,\varepsilon)) \cap \B =\phi \, \,\text{for some}\, \varepsilon>0\}
\label{eq: b_minus}\\
&\t:=\left\{\x \in \b\,\Big|\, \, \F(\x) = \frac{d}{dt}\Big|_{t=0} \boldsymbol \Theta  (\x, t) \,\, \text{is tangent to} \,\b  \right\}
\label{eq:t}
\end{align}

\begin{definition} 
\label{def:iso-block}
If $\t= \b^+ \cap \b^-$ then $\B$ is called to be an \textbf{isolating block}.
\end{definition}

\begin{definition} An isolating block $\B$ \textbf{isolates} $\N$ if int($\B$) is an isolating neighbourhood for $\N$.
\end{definition}

By a theorem of Conley and Easton \cite{CoEa71},  any isolated invariant set accepts an isolating block, and vice-versa, any isolating block contains an isolating block (possible empty).

\medskip
\noindent
A natural way to construct isolated blocks is by using a Lyapunov-type function.

\begin{theorem}[Wilson and Yorke \cite{WilYor73}]\label{lemma:WY} Let $\I:\M \to [0, \infty)$ be a smooth function and $\delta_0>0.$ Suppose
$D\I(\x)\neq 0$ for all $\x$ such that $0<\I(\x)\leq \delta_0$, and whenever $0<\I(\x)\leq \delta_0\,$ and
\[
\frac{d}{dt}\Big|_{t=0} \left( \I(\boldsymbol \Theta (\x, t) \right)) =0
\]
we have,
\[ \frac{d^2}{dt^2}\Bigg|_{t=0} \left( \I(\boldsymbol\Theta (\x, t) \right))>0.\]
Then $\N: = \I^{-1}(0)$ is an isolated invariant set and 
$\I^{-1}\left([0, \delta] \right)$ is an isolating block for $\N$ for each $\delta \in (0, \delta_0].$

\end{theorem}

\medskip
\noindent The subsets of $\b$ asymptotic to $\N$ are 
\begin{align}
&\a^+:= \{\x \in \b^+\,|\,\boldsymbol\Theta (\x, [0, \infty)) \subset \B \}
%= \text{initial conditions in $\b$ that in the past were not in $\b$}
\label{eq: a_plus}\\
&\a^-:= \{\x \in \b^-\,|\, \boldsymbol\Theta (\x, (-\infty, 0]) \subset \B \}\,.
\label{eq: a_minus}
\end{align}
The map across the block $\boldsymbol \Psi: \b^+\setminus \a^+ \to \b^-$ is constructed by assigning to each  $\x \in \b^+ \setminus \a^+ $ the point along the flow $\boldsymbol\Theta (\x, t_{\text{exit}}(\x))$, where 
\[
t_{\text{exit}}(\x):= \text{inf} \{t>0\, |\,\boldsymbol\Theta (\x, t) \notin \B \}
\]
that is
\begin{equation}
\boldsymbol \Psi (\x) = \boldsymbol\Theta(\x, t_{\text{exit}}(\x))\,.
\label{eq:map_across}
\end{equation}
In this context we have
\begin{theorem}\label{IsoBlock}[Conley and Easton \cite{CoEa71}] If $\B$ is an isolating block, then the map across the block $\boldsymbol \Psi: \b^+\setminus \a^+ \to \b^-\setminus \a^-  $ is a diffeomorphism.
\end{theorem} 
\begin{definition} The isolating block $\B$ is  \textbf{trivialisable} if the map across the block $\boldsymbol \Psi$ extends to a diffeomorphism from $\b^+$ to $\b^-.$ 
\end{definition}

\section{Block regularisation of a vector field singularity}
\label{sect:sing_block}

Consider a vector field  \eqref{eq:1} undefined on a compact set $\hat \N \subset \M$ so that trajectories approach $\hat \N$ in finite time. The flow is not complete and for each $\x \in \M$ we denote its orbit ${\cal O}(\x):=\{{\boldsymbol\Theta}(\x, t)\,|\,\boldsymbol\Theta(\x, t) \,\,\text{is well-defined}\}.$ The definitions of $\B$, $\b$, $\b^{\pm}$ from the previous section remain the same, with the understanding that the orbit of points in $\b$ may exist for a finite time only. The same applies for the subsets $\a^{\pm},$ that is
\begin{align}
&\a^+:= \{\x \in \b^+\,|\, \boldsymbol\Theta\left(\x, t \right) \subset \B\,\,\text{for all $t$ for which
 $ \boldsymbol\Theta\left(\x, t\right)$ is defined} \}
%= \text{initial conditions in $\b$ that in the past were not in $\b$}
\label{eq: a_plus_sing}\\
&\a^-:= \{\x \in \b^-\,|\,\boldsymbol\Theta  \left(\x,t\right)\subset \B \,\,\text{for all $t$ for which 
$\boldsymbol\Theta \left(\x, t \right)$ is defined} \}\,.
\label{eq: a_minus_sing}
\end{align}
The map $ {\boldsymbol \Psi} $ is defined in the same manner and the definition of $\B$ being trivialisable remains the same. Finally, 

\begin{definition}\label{def: block}
The vector field \eqref{eq:1}  is \textbf{block regularisable} over the singularity set  $\hat  \N$, or the singularity set $\hat \N$ is  block regularisable,  if there is a trivialisable  block that isolates $\hat  \N$. 
\end{definition}

\medskip

Assume that there exists a (at least) ${\cal C}^2$ transformation  of the variables and a time-reparametrisation,  so that when applied to \eqref{eq:1}  the set $\hat \N$ is blown up (or transformed) into a set $\N$ pasted into the  phase space that is approached by the (re-parametrized) flow asymptotically and for which $\N$  is a compact invariant manifold. 
%
%
%In our case, as we will see shortly, $\hat \N = \{(x,y,p_x, p_y)\in \mathbb{R}^4\,|\,  x=y=0\}$ is blown up to a set $\N =\{r, \theta, p_r, \p_\theta) \in [0, \infty) \times {\cal S}^1 \times \mathbb{R}^2 \,|\, r=0\}$; intuitively, it consists in the blown up of the origin $r=0$ of the plane, where polar coordinates are not defined, with "a point $\times$ a direction".
%
%
%the flow becomes complete, that is all trajectories that approached $\N$ in finite time now approach it asymptotically, and $\N$ is  a compact invariant manifold  pasted into the  phase space. 
We say then that  the equations of motion are \textit{regularised}.
%
% and 
%$\N$ is  \textit{blown out} into a compact invariant manifold $\N$ pasted into the  phase space. In the new re-parametrisation all trajectories that approached $\hat \N$ in finite time now approach $\N$ asymptotically and the flow is complete, then the equations of motion are called to be \textit{regularised}.
Notice that the flow on invariant set $\N$ is fictitious, but due to continuity with respect to initial data, it provides information on the behaviour of the real flow near $\hat \N$. 

If the blow-out procedure is applicable, then the definition \ref{def: block} above is equivalent to

\begin{definition}
The vector field \eqref{eq:1}  is \textbf{block regularisable} over the singularity set  $\hat \N$ if there is a trivialisable block that isolates the blown-up invariant manifold $\N$.
\end{definition}

\section{The logarithm central force problem. Regularisation of the equations}
\label{sect:ln}

\noindent
Consider the motion of a unit mass particle described by the Hamiltonian $H: \mathbb{R}^2\setminus\{{\bf 0}\} \times \mathbb{R}^2 \to \mathbb{R}$
\begin{align}
H(\q, \p) =  \frac{1}{2}\p^2 +\ln(|\q|)
\end{align}
The equations of motion are
\begin{align}
&\dot \q  = \p\,, \quad \quad \dot \p = -\frac{\q}{|\q|^2}\,. \label{eq:init_H_eq}
\end{align}
In polar coordinates, the Hamiltonian reads
\begin{align}
H(r, \theta, p_r, p_{\theta}) =  \frac{1}{2}\left(p_r^2 + \frac{{p_\theta^2}}{r^2} \right) +\ln(r)
\label{E: Ham_2}
\end{align}
and the equations of motion are
\begin{align}
&\dot r  = p_r \,,\quad \,\,\,\, \,\,\,\,\,\dot p_r =  \frac{p_\theta^2}{r^3}-\frac{1}{r}\,, \label{eq:init_1} \\
&\dot \theta = \frac{p_\theta}{r^2}\,,  \quad \,\,\,\,\,\,\,\dot p_\theta = 0\,.  \label{eq:init_2}
\end{align}
By physics laws, or by direct observation, one may deduce that along any solution $\left( r(t), \theta(t), p_r(t), p_{\theta}(t) \right)$ the total energy $H(r(t), \theta(t), p_r(t), p_{\theta}(t))$ and the  angular momentum $p_\theta(t)$ are conserved and thus can be considered as parameter. Denoting  $p_\theta=: c$, we reduce the dynamics to a one degree of freedom system given by the reduced (or amended) Hamiltonian
\begin{align}
H_{\text{red}}(r, p_r) =  \frac{1}{2}p_r^2 + \frac{c^2}{2r^2} +\ln(r)
\label{eq: init_Ham}
\end{align}
Since along any motion we have $H_{\text{red}}(r(t), p_r(t)) = const.=:h$, the Hill regions of motion are given by inequality
\begin{align}
V_{\text{red}}(r): = \frac{c^2}{2r^2} +\ln(r) \leq h
\end{align}
It follows that collision are possible only for zero angular momentum (i.e. for $c=0$), and that for any $c\neq 0$ the motion is possible for $h \geq h_{min}(c) = 1/2 + \ln(|c|)$. For all levels of energy the motion  is always bounded, with the bounds depending on the parameters $(h,c)$, that is $r(t)\in [r_{min}(h,c), r_{max}(h,c)]\,.$

\bigskip
\noindent
The equations of motion associated to \eqref{eq: init_Ham} are
\begin{align}
&\frac{dr}{dt} =  p_r \label{eq: eq_init_Ham_1}\\
&\frac{dp_r}{dt} = \frac{c^2}{r^3}-\frac{1}{r}
\label{eq: eq_init_Ham_2}
\end{align}
By the standard ODE theory, for any initial conditions $\x(0)$ there is a unique solution defined on a maximal interval of existence $(t^-, t^+)$, with $-\infty\leq t^- <0<t^+\leq \infty$. Given that the system is time-reversible, $|t^-| = t^+$, and so the solution exists for $t \in (-t^+, t^+).$ If $t^+< \infty$ , then the solution experience a \textit{singularity} at  $t^*=t^+$.  
In our case, a proof similar to that in \cite{McGe81} shows that if solution experience a singularity at $t^*$, then $\displaystyle{\lim \limits_{t \to t^*}r(t)\to 0}$.

%for initial conditions with $c=0$, solutions of the system \eqref{eq: eq_init_Ham_1}-\eqref{eq: eq_init_Ham_2} have singularities and are defined for $t\in (-t^*, t^*)$ for some $0<t^*< \infty.$

\bigskip
\noindent
We proceed now to regularize the equations. First, in \eqref{eq:init_H_eq}  we apply the change the variables $(\q, \p)\to (\s, \u)$:
\begin{align}
\q = re^{-\frac{1}{r^2}}\s\,,\,\,\,\,\,\,\p =\frac{1}{r}\u
\label{eq: transf_1}
\end{align}
where $(r, \s, \u) \in (0, \infty) \times S^1 \times \mathbb{R}^2$. In these coordinates the conservation of energy reads
\begin{align}
\frac{1}{2}|\u|^2 + r^2 \ln r -1 =h r^2\,.
\label{eq:energy}
\end{align}
 The system \eqref{eq: transf_1} is analytic on its domain and that the regions of motion are constrained by the energy relation \eqref{eq:energy}. Notice that the kinetic term in \eqref{eq:energy} is positive, we have that for a fixed level of energy h
 \[
 hr^2- r^2 \ln r+1\geq 0.
 \]
It follows that  for any fixed level of energy $h \in \mathbb{R}$, the motion exists  and the trajectory is bounded, i.e., $0 \leq r\leq r_{\text{Max}}(h).$
 We now express the coordinates $(\s, \u)$ in terms of  $(w,\phi, \psi) \in \R\times S^1 \times S^1$:
\begin{align}
&\s =(\cos \phi, \sin \phi)\,,\quad \quad \u= e^w (\cos \psi, \sin \psi)\,.
\end{align}
 Defining further $\varphi:=\phi-\psi$ and introducing the time-reparametrisation

\begin{align}
d \tau =- \frac{1}{r^2}\exp \left(\frac{1}{r^2} - w\right) dt
\label{eq:time_reparam} 
\end{align}
the equations of motion become %
\begin{align}
&\dot r  = -\frac{r^3}{r^2+2} e^{2w}\cos\varphi 
\label{E:1}\\
&\dot \varphi= (e^{2w} -r^2)\sin \varphi\,,
\label{E:2} \\
&\dot \psi=r^2 \sin\varphi\,,
\label{E:3} \\
&\dot w = r^2\left[1 - \frac{e^{2w}}{r^2+2}\right]\cos \varphi \,\,\,\, \text{with}\,\,\,\,\,(r, \varphi, \psi, w)\in (0, \infty) \times S^1 \times S^1 \times \R
\label{E:4}
\end{align}
(where, by abuse of notation, we denote  the differentiation with respect to $\tau$ by ``dot" as well). Notice that the vector field has a smooth extension to $r = 0$. To extend the energy relation $\eqref{eq:energy}$, we define
\[
f(r) =\begin{cases}
    r^2\ln r\,,\,\,\,r>0\,, \\
     0\,,\,\,r=0,
\end{cases}
\]
and observe that $f$ is of ${\cal C}^1$ class. This yields, %We extend the vector field \eqref{Eq:1}-\eqref{Eq:3} and the energy relation \eqref{eq:energy} to include points with $r=0$ using $f(r)$ defined above:
%
%\begin{align}
%&\dot r  = -\frac{r^3}{r^2+2} 2\left(hr^2 - f(r) +1 \right)\cos\varphi\,, 
%\label{E:1}\\
%&\dot \varphi= \left[2\left(hr^2 - f(r) +1 \right) -r^2\right]\sin \varphi\,,
%\label{E:2} \\

%&\dot \psi=r^2 \sin\varphi\,,\,\,\,\, \text{with}\,\,\,\,\,(r, \varphi, \psi)\in[0, \infty) \times S^1 \times S^1
%\label{E:3}
%\end{align}
%
%
%
%
%
 \begin{equation}
 \frac{1}{2}e^{2w}=hr^2- f(r)+1\,.
\label{EEO} 
 \end{equation}

Notice that  the extended flow \eqref{E:1}-\eqref{E:4}  is complete; we denote it by $\theta_t$. \medskip
Further, notice that $\psi$ is determined by $r$ and $\varphi$ and, for $r$ small but positive, since  
$e^{2w} - r^2 = \left[(2h - 1 - \ln r)r^2 +1 \right] \geq 0$ 

\begin{itemize}
\item if $\varphi \in (-\pi/2, \pi/2)$ then $r$ is decreasing, and furthermore, if $\varphi \in (-\pi/2, 0)$ then $\varphi$ is decreasing, whereas if $\varphi \in (0, \pi/2)$ then $\varphi$ is increasing

 \item if $\varphi \in (\pi/2, 3\pi/2)$ then $r$ is decreasing, and furthermore, if $\varphi \in (\pi/2, \pi)$ then $\varphi$ is increasing, whereas if $\varphi \in (\pi, 3\pi/2)$ then $\varphi$ is decreasing.
\end{itemize}

\noindent
In $(r, \varphi, \psi, w)$ coordinates, the angular momentum integral is
\begin{align}
\exp \left(w-\frac{1}{r^2} \right) \sin \varphi =const.=: c
\label{int: ang_mom}
\end{align}
and it is undefined  at $r=0.$ By introducing 
\begin{align}
g(r) =  
\begin{cases}
\exp \left(-\frac{1}{r^2} \right)\,,\,\,\,r>0  \\
 0\,,\,\, r=0\,.
\end{cases}
\end{align}
we extend  \eqref{int: ang_mom} to a  well-defined and differentiable function, including points with $r=0$. The extended angular momentum relation reads:
\begin{align}
e^wg(r) \sin \varphi =const.=: c\,.
\end{align}

\section{Block regularisation of the logarithm problem}

As mentioned above, the flow of  \eqref{E:1}-\eqref{E:4}  is complete. For this flow, the collision set $\{r=0\}$ was blown up to 
the compact set
\begin{align}
{\N}:&=
\{(r, \varphi, \psi, w)\in [0, \infty) \times S^1 \times S^1 \times \R|\, r=0, e^{2w} = 2\}
\end{align}
which is an invariant manifold, called the \textit{collision manifold}, pasted into the phase space for all levels of energy. The flow on $\N$ is fictitious and has no physical meaning, however, by the continuity with respect to initial data, one may extract information on the real flow behaviour near collision. On $\N$ equation  of motion are
\begin{align}
&\dot \varphi= 2 \sin \varphi\,,
\label{EC:2} \\
&\dot \psi=0\,.
\label{EC:3}
\end{align}
and so, on  $\N$ the flow is trivial,  given by trajectories with $\psi (\tau)=const.=\psi_0.$ There are two circles of equilibria at  $(\varphi, \psi )= (0,\psi_0)$ and $(\varphi, \psi) = (\pi,\psi_0)$ for all $\psi_0 \in S^1$. The next lemma shows that the collision manifold is approached asymptotically in time.

\begin{lemma}\label{Lemma_1} Let $h$ be fixed and 
let $(r, \varphi, \psi, w)$  be a solution of \eqref{E:1}-\eqref{E:4}. Assume that $r \to 0$ as $t \to t^{*}\pm.$ Then $\tau(t) \to \pm \infty$.
\end{lemma}

\noindent
Proof. We prove that $t\to t^{*}-$ implies $\tau(t) \to -\infty;$ the other case is analogous.  The re-parametrisation \eqref{eq:time_reparam}  yields
\begin{align}
\tau(t) = \tau_0 - \int\limits_{t_0}^t \exp \left(\frac{1}{r^2(\sigma)} - w(\sigma)\right)  \frac{1}{r^2(\sigma)}d\sigma\,.
\nonumber
\end{align}
Thus $\tau(t)$ is defined for all $t < t^-$ and is decreasing. Thus $\lim\limits_{t \to t^{*}-} \tau(t) =  \tau^{*}\geq -\infty.$ Then $\lim\limits_{\tau \to \tau^{*}} r(\tau) = 0.$ But equations \eqref{E:1}-\eqref{E:4} are well-defined and with a smooth vector field. Since $\N$ (where $r=0$) is a compact invariant set, trajectories cannot approach it in finite time. Thus $\tau^{*} = -\infty.$

\noindent $\square$

\bigskip
Define the sets
  \begin{align}
%{\bf S}^{\pm}:=\{(r, \varphi, \psi, w)\in [0, \infty) \times S^1 \times S^1 |\,r=0\,, u=\pm \sqrt{2}\}
{\bf S}^{+}:&=\{(r, \varphi, \psi, w)\in [0, \infty)\times S^1 \times S^1 \times \R |\,r=0\,, \varphi = 0\,, e^{2w} = 2\}\\
{\bf S}^{-}:&=\{(r, \varphi, \psi, w)\in [0, \infty) \times S^1 \times S^1 \times \R|\,r=0\,, \varphi = \pi\,,e^{2w} = 2\}
\end{align}
and denote by $\boldsymbol{\omega}(P)$ the omega limit set of a point $P=(r, \varphi, \psi, w)$ under the flow of \eqref{E:1}-\eqref{E:3} on a fixed level of $h$.
 %For $h$ fixed, in the $(r, \varphi, \psi)$ phase-space, the points
 %
 % \begin{align}
%E^-:=\{(0, 0, \psi_0\,|\,\psi_0 \in S^1\}\,,\,\,\,E^+:=\{(0, \pi, \psi_0\,|\,\psi_0 \in S^1\}
%\end{align}
%
%are non-isolated (degenerate) equilibria, with $E^-$ being a sink, and $E^+$  a source (along the $r$-direction).

\begin{lemma}\label{Lemma_2}
For $h$ fixed, consider a solution $(r(\tau), \varphi(\tau), \psi(\tau), w(\tau))$ of the  \eqref{E:1}-\eqref{E:4}, with some initial condition $P_0:=(r(0), \varphi(0), \psi(0), w(0))$. Then $r(\tau)\to 0$ as $\tau \to \infty$ if and only if $\boldsymbol{\omega}(P_0)={\bf S}^{+}$, and  $r(\tau)\to 0$ as $\tau \to -\infty$  if and only if $\boldsymbol{\alpha}(P_0)={\bf S}^{-}.$
\end{lemma}
Proof: We will only prove that $r(\tau)\to 0$ as $\tau \to \infty$ if and only if $\boldsymbol{\omega}(P_0)={\bf S}^+$ since the other one is similar. Let $\theta_\tau (P_0)$ denote the trajectory starting at $P_0$. Sufficiency follows from that fact that if  $\boldsymbol{\omega}(P_0)={\bf S}^{+}$ then, by definition, $\theta_\tau (P_0) \rightarrow y, y \in \bf S^{+}$. But this implies that $r_{\tau} \rightarrow 0$.

To prove the converse assume that $r(\tau) \rightarrow 0$. Then $\bf S^+ \subseteq \boldsymbol{\omega} (P_0)$, so we need to prove the reverse inclusion. From the continuity of the angular momentum equation, if $r \rightarrow 0$, we conclude that $c = 0$. Thus, any trajectory going to $\mathbf{N}$ must have angular momentum 0. For such a trajectory, we must have $\sin \varphi = 0$ from the angular momentum equation. By continuity, this means that $\varphi(0) \in \{0, \pi\}$ and hence, $\varphi$ remains constant for all $\tau \in [0,\infty)$. If $\varphi(\tau) = \pi$ for all $\tau \geq 0$ then in a sufficiently small neighborhood of N, $r$ increases and hence $r(\tau) \nrightarrow 0$ as $\tau \rightarrow \infty$. We conclude that $\varphi(0) = \varphi = 0$. Since, by assumption, $r(\tau) \rightarrow 0$, we conclude that $\boldsymbol\omega (P_0) \subseteq \bf S^+$.

\noindent $\square$

\begin{lemma}\label{Lemma_3}
Let $(r, \varphi, \psi,w)$ a solution of \eqref{E:1}-\eqref{E:4} such that $r \to 0$ as $\tau \to \pm \infty.$ Then, as $\tau\to \pm \infty$, 
\begin{align}
c \equiv 0\,,\,\,\,\,\sin \varphi (\tau) \equiv 0\,,\,\,\,\,\psi(\tau) = const. =  \psi^*   \,\,\,\,\text{and}\,\,\,\, \, e^{2w(\tau)} \to 2\,.
\end{align}
\end{lemma}

\noindent
%Since from the previous lemma, for a solution $(r(\tau), \varphi(\tau), \psi(\tau), w(\tau))$ with $(r(0), \varphi(0), \psi(0), u(0)) =:P_0$ we have that 
%$r \to 0$ as $\tau \to \pm \infty$ if and only if $\boldsymbol{\omega}(P_0)={\bf S}^{-}$, we need to consider points $P$ on the stable manifold of ${\bf S}^{-}$ and investigate what %are the limits of $\varphi (\tau), \psi(\tau)
%$ and $u(\tau)$ as $\tau \to \infty.$

%Consider the stable manifold of ${\bf S}^+$ and fix $h.$ Further, let $P$ be on the 

%By the previous Lemma, it is sufficient to prove that the limits of all points on  fixed level of energy $h$ on the stable manifold of ${\bf S}^+$ are as above.

 %ll From the energy integral \eqref{EEO} it is clear  that $u \to \pm \sqrt{2}.$ As $r \to 0$, we have that $\dot \varphi \to 0$ and so $\varphi(t) \to $

\noindent Proof. Following a similar analysis as in the previous proof, we conclude that $r \to 0$ as $\tau \to \pm \infty$ implies $c\equiv 0$, and consequently $\sin \varphi (\tau) \equiv 0$. The remaining conclusions follow by using equation \eqref{E:3} with $\sin\varphi = 0$ and the energy equation.

\noindent $\square$

\bigskip

\bigskip

\noindent For a given $h$, using the conservation of energy equation, \ref{EEO}, we define \[\mathbf{\Sigma}_h := \{(r,\varphi,\psi, w) \in [0,\infty)\times S^1 \times S^1 \times \R |\hspace{5pt} (\ref{EEO}) \text{ holds}\}.\]
It follows that the flow of \eqref{E:1} - \eqref{E:4} is a map $\mathbf{\Sigma}_h \times \mathbb{R} \rightarrow \mathbb{R}$. Define $\I:\mathbf{\Sigma_h} \rightarrow [0, \infty)$ by $\I(r,\varphi,\psi,w) = r$. The tangent space to a point $(r,\varphi,\psi,w) \in \mathbf{\Sigma}_h$ is \[\{(\dot{r},\dot{\varphi},\dot{\psi}, \dot{w})| e^{2w}\dot{w} + (f'(r) - 2hr)\dot{r} = 0\}.\] Note that, for a fixed $h$, $r$ can be chosen small enough and positive so that $f'(r) - 2hr = (2 \ln r + 1 -2h)r \neq 0$. This means that, there exists $\delta >0$ such that if $0 < r < \delta$ then $D\I(r,\varphi,\psi, w) = \dot{r} \neq 0$.
\medskip
\noindent Next, assume that
 \[
 \dot\I(r,\varphi,\psi,w):=\frac{d}{dt}\Big|_{t=0}\I(r(t),\varphi(t),\psi(t),u(t))= \dot r (0) = 0
 \] 
 and $\I(r(0),\varphi(0),\psi(0),u(0))=r(0)=: r_0 > 0$. From  \eqref{E:1}, for $0 <r_0 < \delta$ we must have $\cos \varphi(0) = 0$. Then
%Choosing $r < \min \delta = \{u^2, \delta_1\}$, we have

\[
    \frac{d^2}{dt^2}\Big|_{t=0} \I (r(t),\varphi(t),\psi(t),w(t)) = \ddot{r}(0)= \frac{r^3_0}{r^2_0 + 2}\left(e^{2w(0)} - r^2_0\right)>0
\]
assuming $\delta$ is sufficiently small. 

%\textbf{I REWROTE THIS PART, CAN YOU PLEASE CHECK IT ONCE IF EVERYTHING IS OKAY?}

\bigskip

\noindent Let
\begin{equation}\label{block}
    \B := \{(r,\varphi,\psi,w)|r \leq \delta\}.
\end{equation}
The following proposition is a consequence of the Wilson and Yorke Lemma \ref{lemma:WY} and our prior discussion:

\begin{proposition}
$\B$ is an isolating block for the isolated invariant set \bf{N}.
\end{proposition}

\medskip

\begin{figure}[h!]
\centering
  %     \subfigure[ Cases $C>0,$ $D>0$ for $2C-3D<0$ (top), $2C-3D=0$ (middle), $2C-3D>0$ (bottom)  ]
      %{
       %    \includegraphics[angle=0,scale=0.3] {C>0-and-D>0.jpg}
        %   }
           \includegraphics[angle=0,scale=0.6] {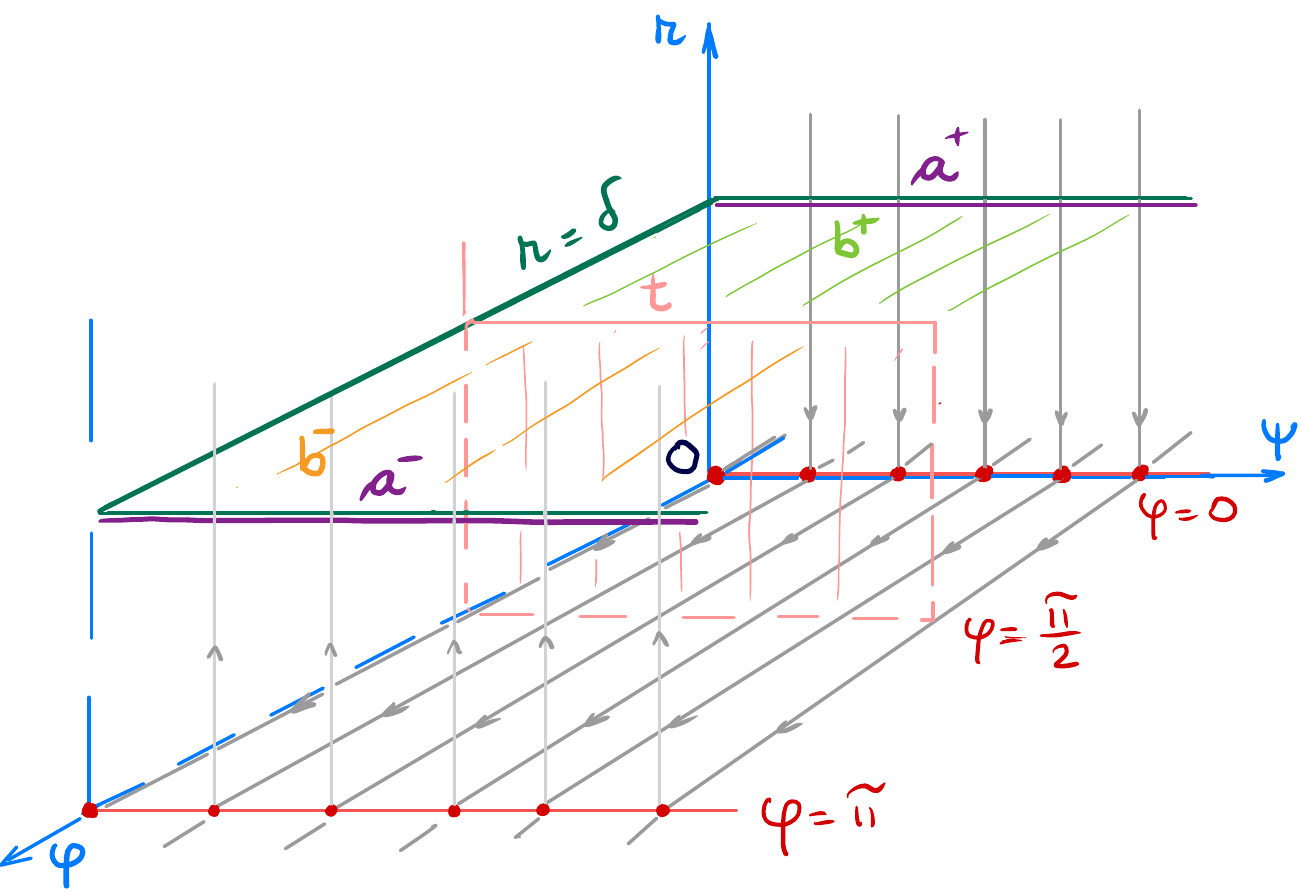}
           \caption{The sets $\b^{\pm}$, $\a^{\pm}$, $\t$, and the flow on the collision manifold.} 
         \label{figure}
\end{figure}

\noindent
Denote by $w_{\delta}$ the solution of $\eqref{EEO}$ with $r = \delta$. We let
\begin{align}
\b=\partial \B  = \{((r, \varphi, \psi,w)\,|\, r= \delta \,,(\varphi, \psi)\in S^1 \times S^1 \,, w = w_{\delta}\}\,.
\end{align}
\noindent 
and we define, (see Figure \ref{figure})
\begin{align}
&\b^+:=\{(r, \varphi, \psi,w)\,|\,r=\delta\,,\cos \varphi \geq 0, \psi \in S^1, w = w_{\delta}\} \,,\,\,\,\,\b^-:=\{(r, \varphi, \psi,w)\,|\,r=\delta\,,\cos \varphi \leq 0\,, \psi \in S^1, w = w_{\delta}\}  \\
&\t:=\{(r, \varphi, \psi,w)\,|\,r=\delta\,,\cos \varphi = 0, w = w_{\delta}\}  = \{(r, \varphi, \psi, w)\,|\,r=\delta\,, \varphi = \pi/2\,\,\text{or}\,\,\,\varphi=3\pi/2 , w = w_{\delta}\} \\
&\a^+:=\{(r, \varphi, \psi,w)\,|\,r=\delta\,,  \varphi = 0, w = w_{\delta}\} \,,\,\,\,\a^-:=\{(r, \varphi, \psi,w)\,|\,r=\delta\,,  \varphi = \pi, w = w_{\delta}\} 
\end{align}

Denoting by $\boldsymbol \Phi_t(r, \varphi, \psi,w)$ the flow of \eqref{E:1}-\eqref{E:4}, notice that the definitions above are in agreement with the definitions  \eqref{eq: b_plus} - \eqref{eq: a_minus}. Indeed, if $\cos \varphi(0) \leq 0$ and $r(0) = \delta$, then $r$ increases in $(0,\epsilon)$ for sufficiently small $\epsilon > 0$. In that case, $\boldsymbol \Phi_{(0,\epsilon)}(\delta,\varphi,\psi,w_{\delta}) \cap \bf B = \emptyset$ and therefore, 
\[
\b^- = \{(r,\varphi,\psi,w) \in \b\, | \,\boldsymbol \Phi_{(0,\epsilon)}(\delta,\varphi,\psi,w_{\delta}) \cap \bf B = \emptyset \text{\: for some\:} \epsilon> 0\}
\]
where equality holds since $\cos \varphi(0) > 0$ implies $r$ decreases and hence for every $\epsilon > 0$, $\Phi_{(0,\epsilon)}(\delta,\varphi,\psi,w_{\delta}) \cap \bf B \neq \emptyset$. In a similar vein, 
\[
\b^+ = \{(r,\varphi,\psi,w) \in \b\, |\, \boldsymbol \Phi_{(-\epsilon,0)}(\delta,\varphi,\psi,w_{\delta}) \cap \bf B = \emptyset \text{\: for some\:} \epsilon> 0\}
\]
and  $\t  = \{(r,\varphi,\psi,w)\, |\, \dot{\boldsymbol \Phi}_0(r,\varphi,\psi,w) \text{\: is tangent to \:}\b \}$ by definition of an isolating block. 
\bigskip

Next, if $\varphi = 0$ along the flow then from \eqref{E:1}-\eqref{E:4}, it follows by substituting $\varphi = 0$ that $\dot{r} < 0$. So, if $r(0) = \delta$, then the flow stays inside the block. Hence, we have, $\a^+ \subseteq \{(r,\varphi,\psi,w)\in \b^+| \boldsymbol \Phi_{[0,\infty)}(r,\varphi,\psi,w) \subseteq \B\}$. On the other hand, the $\boldsymbol{\omega}$-limit set of a point $(\delta, \varphi_0,\psi_0,w_\delta) \in\{(r,\varphi,\psi,w)\in \b^+| \boldsymbol \Phi_{[0,\infty)}(r,\varphi,\psi,w) \subseteq \B\}$ is in the collision manifold $\N$. This means that $\sin\varphi_0 = 0$ and $\boldsymbol{\omega}(\delta, \varphi_0,\psi_0,w_\delta) = \S^+$ which yields $\varphi_0 = 0$. Thus, $\a^+ = \{(r,\varphi,\psi,w)\in \b^+\,|\, \boldsymbol \Phi_{[0,\infty)}(r,\varphi,\psi,w) \subseteq \B\}$ and similarly, $\a^- = \{(r,\varphi,\psi,w)\in \b^-\,|\, \boldsymbol \Phi_{[-\infty,0)}(r,\varphi,\psi,w) \subseteq \B\}.$    

\bigskip
Now we define the map across the block
\begin{align}
\Psi: &\b^+ - \a^+  \longrightarrow \b^-  \nonumber \\
&(\delta, \varphi_0, \psi_0,w_{\delta})  \longrightarrow \Psi(\delta, \varphi_0, \psi_0,w_{\delta}) := 
 \boldsymbol \Phi_{t_{exit}}(\delta, \varphi_0, \psi_0,w_{\delta}) \\
&\hspace{4.4cm}\, =\left(\delta, \varphi(t_{exit}; (\delta, \varphi_0, \psi_0), \psi(t_{exit}; (\delta, \varphi_0, \psi_0),w_{\delta}\right)  
=:(\delta, \varphi_{exit}, \psi_{exit},w_{\delta}) \,,
\nonumber
\label{Def: Psi}
\end{align}
where $t_{exit} = $ the time of exit from $\B$. With this definition, Theorem \ref{IsoBlock} holds.

\medskip
Consider the flow starting from a point $\mathbf{P}_0 = (\delta,0,\psi_0,w_{\delta}) \in \mathbf{a}^+$. As $r \rightarrow 0$, by Lemma $\ref{Lemma_2}$, $\boldsymbol{\omega (P_0)}$ is the point $\mathbf{s}^+ = (0,0,\psi_0,(1/2)\ln 2) \in \mathbf{S}^+$. The flow emerging from $\mathbf{s}^+$ then follows the unstable manifold of $\mathbf{s}^+$. Since the unstable manifold on $\mathbf{s}^+$ lies in the collision manifold $\mathbf{N}$, we proceed to study the flow on $\mathbf{N}$. 

\medskip
As remarked before, on $\mathbf{N}$ the trajectories are given by $\psi(\tau) = \text{constant}$. Thus, the flow follows the unstable manifold of $\mathbf{s}^+$ to a point $\mathbf{s}^- = (0,\pi,\psi_0, (1/2)\ln 2)\in \mathbf{S}^-$. But from Lemma $\ref{Lemma_2}$, the flow follows the unstable manifold of $\mathbf{s}^-$ till it reaches the point $(\delta,\pi,\psi_0,w_{\delta}) \in \mathbf{a}^-$. 

%With this in mind, we define
%
 %\[\Psi(\delta, 0 , \psi_0) = (\delta, \pi,\psi_0).????\]

\begin{proposition}
In the logarithm  problem, the map across the block extends to a homeomorphism from $\b^+$ to $\b^-$. 
\end{proposition}

\noindent
Proof: At $r=\delta,$ the angular momentum integral reads
\[2\exp(w_{\delta}-1/\delta^2)\sin \varphi=c\]
For $c>0$ we have $\varphi (0) \in (0, \pi)$ and at the time of exit, since $r(t_{exit})=\delta$, we must have $\varphi(t_{exit})=\varphi (0)$ or $\varphi(t_{exit})= \pi-\varphi (0).$ But from \eqref{E:2}, $\varphi$ is increasing and so we must have $ \varphi(t_{exit})= \pi-\varphi(0).$ Further, from \eqref{E:3}, with $\varphi(0) = \varphi_0$, we can write,
\begin{align}
\psi(t_{exit}) =  \psi(0) + \int\limits_0^{t_{exit}} r^2(t; \delta) \sin \varphi(t; \varphi_0)dt
\end{align}
Since $r(t)$, $\varphi(t)$ and $u(t)$ so not depend on $\psi$, and their evolution is determined by their initial conditions $(\delta, \varphi_0, w_{\delta})$, and since $\delta$ is fixed, we have $ \int\limits_0^{t_{exit}} r^2(t; \delta) \sin \varphi(t; \varphi_0)dt =  G(\varphi_0)$, where $G$ is some function. 
Consequently, for $c>0$ (and thus $\varphi_0 \in (0, \pi)$) the map $\Psi$ reads
\begin{align}
\Psi(\delta, \varphi_0, \psi_0, w_{\delta})  = (\delta, \pi-\varphi, \psi_0 + G(\varphi_0),w_{\delta})
\label{E:c>0}
\end{align}
Similarly, for $c<0$ (and thus $\varphi_0 \in (\pi, 2\pi)$) the map $\Psi$ reads
\begin{align}
\Psi(\delta, \varphi_0, \psi_0,w_{\delta})  = (\delta, \pi+\varphi_0, \psi_0 + G(\varphi_0),w_{\delta})
\label{E:c<0}
\end{align}
Define $\Psi(\delta,0,\psi_0,w_{\delta}) = (\delta,\pi,\psi_0,w_{\delta})$. Since the system is time-reversible, it suffices to show that there is continuous extension of $\Psi$ from $\b^+$ to $\b^-$. In order to prove that $\Psi$ extends to a continuous map $\b ^+ \rightarrow \b ^-$, we need to show that $\lim_{\varphi_0 \rightarrow 0}G(\varphi_0) = 0$. First note that, $\lim_{\varphi_0\rightarrow 0}r^2(t;\delta)\sin \varphi(t;\varphi_0) = r^2(t;\delta)\sin \varphi(t;0)$, since the vector field in \eqref{E:1} - \eqref{E:4} is smooth. This means that $(r(0),\varphi(0),\psi(0), u(0)) = (\delta,0,\psi_0,w_{\delta}) \in \a ^+$. But then, $r \rightarrow 0$ as $\tau \rightarrow \infty$. This means, by Lemma \ref{Lemma_3}, that $\sin \varphi \equiv 0$. Hence, $\lim_{\varphi_0\rightarrow 0}r^2(t;\delta)\sin \varphi(t;\varphi_0) = 0$. Note that since $r \leq \delta$, it follows that $|G(\varphi_0)| \leq 2 \delta^2$. Thus, by using the Lebesgue Dominated Convergence Theorem and noting that since the flow starting from a point in $\a ^+$ remains in the block, the time of exit for such a point is infinity, we obtain,

\[\lim_{\varphi_0 \rightarrow 0} G(\varphi_0) = \int_0^{\infty}\lim_{\varphi_0 \rightarrow 0}r^2(t; \delta) \sin \varphi(t; \varphi_0)dt = 0.\]

\noindent which completes the proof.

\noindent
$\square$

\begin{remark}
\label{remark}
In the case of homogeneous potentials of the form
\begin{equation}
V(r)=-\frac{1}{r^{2-2/n}}\,,\,\,\,n\geq 2
\label{eq: reg_pot}
\end{equation} 
 the map across the block is extended to a diffeomorphism  by removing the sets $\a ^+$ and $\a ^-$ through the use of a generalisation of the Levi-Civita transformation and rescaling the energy  \cite{McGe81}. Specifically, identifying the plane of the motion with the complex plane, this is achieved with the help of the conformal transformation $z \mapsto z^n$. (For more on conformal transformations applied to planar central force problems,  see \cite{GBM08}.) Our attempts in finding a regularising conformal transformation in the case of the logarithm potential were not successful; this problem remains open. 
%
%
%
%see also\cite{GBM08}. Furthermore, in \cite{GBM08}, it is shown that conformal transformations of planar central force problems lead to dual central force motions in the complex plane under an appropriate time reparametrization, which, in the case of McGehee potentials, effectively regularizes the problem by transforming the equations to equations which do not singularities. The problem of finding an appropriate conformal transformation for the logarithmic problem in order to extend the map across the block to a diffeomorphism remains as a prospect for future research.
\end{remark}

\section{Acknowledgements} This work was partially supported by a NSERC Discovery Grant.


\begin{thebibliography}{99}




%A. B\'erard Y. Grandati and H. Mohrbach. Complex representation of planar motions and conserved quantities of the Kepler and Hooke problems. Journal of Nonlinear Mathematical Physics, 17(2):213–225, 2010.

    
 \bibitem[BBP07]{BBP07}  C. Belmonte, D. Boccaletti, and G. Pucacco, On the orbit structure of the logarithm potential, The Astrophysical Journal \textbf{669} (2007), 202-217  
   



\bibitem[BiTr87]{BiTr87}J. Binney and  S. Tremaine, Galactic Dynamics, Princeton,NJ, Princeton University Press, 1987

\bibitem[CaTe11]{CaTe11}R. Castelli and  S. Terracini, On the regularization of the collision solutions of the one-center problem with weak forces, Discrete and Continuous Dynamical Systems (DCDS), \textbf{31} (2011),1197-1218


\bibitem [ChHs22]{ChHs22}JKuo-Chang Chen and Ku-Jung Hsu
The collision singularity of the Kepler problem with singular perturbations, Proceedings AMS, in press
DOI: https://doi.org/10.1090/proc/15600 

\bibitem[CoEa71]{CoEa71}
C. Conley and R. Easton, Isolated invariant sets and isolating blocks, Trans. Amer. Math. Soc, 
\textbf{158} (1971), 35-61


\bibitem[GBM08]{GBM08} Y. Grandati, A. Berard, H. Mohrbach, Complex representation of planar motions and conserved quantities of the Kepler  and Hooke problems, Journal of Nonlinear Matyematical Physics \textbf{17} (2010), 213-225 
% https://hal.archives-ouvertes.fr/hal-00264946v1

\bibitem[HedL12]{HedL12} G. Heckman and T. de Laat. On the regularization of the Kepler problem. Journal of Symplectic Geometry, 10:463–473, 09 2012.


 \bibitem[KS65]{KS65}P. Kustaanheimo and E. Stiefel. Perturbation theory of Kepler motion based on spinor regularization. J. Reine Angew. Math. \textbf{218} (1965) 204-219



\bibitem[Landau60]{Landau60} L.D. Landau and E.M. Lifshits, Mechanics, Pergamon Press, Oxford-London-Paris, 1960

\bibitem[LC13]{LC13}T. Levi-Civita, Nuovo sistema canonico di elementi ellittici. Ann. Mat. Pura Appl., \textbf{20}(1913), 153–169



\bibitem[McGe81]{McGe81}
R. McGehee, Double collisions for a classical particle system with non-gravitational interaction
Comment. Math. Helvetici \textbf{56} (1981) 524-557

\bibitem[Mil83]{Mil83}John Milnor. On the geometry of the Kepler problem. The American Mathematical Monthly, 90(6) (1983), 353–365.

\bibitem[MESc89]{MESc89}J. Miralda-Escude, M Schwarzschild, On the orbit structure of the logarithmic potential, The Astrophysical Journal \textbf{339} (1989),  752-762

\bibitem[Mo70]{Mo70}J. Moser, Regularization of Kepler's problem and the averaging method on a manifold, Communications in Pure and Applied Mathematics \textbf{23} (1970) 609-636


\bibitem[San09]{San09} M Santoprete, Block regularization of the Kepler problem on surfaces of revolution with positive constant curvature, Journal of Differential Equations \textbf{247} (2009)1043-1063

%\bibitem[Sc79]{Sc79} M. Schwarzschild, 1979Astrophys.J.232236

\bibitem[St00]{St00}
C. Stoica, Particle systems with quasihomogeneous interactions, PhD Thesis, University of Victoria, Canada, 2000 


\bibitem[StSc71]{StSc71}E.L. Stiefel  and G. Scheifele, Linear and Regular Celestial Mechanics, Springer and Verlag, Berlin, 1971

%\bibitem Stoica, Gheorghe; Stoica, Cristina; Mioc, Vasile Branch regularization of quasihomogeneous %functions. Rev. Roumaine Math. Pures Appl. 45 (2000), no. 5, 897–905 (2001)

\bibitem[St02]{St02}  C. Stoica, Classical scattering and block regularization for the homogeneous central field problem, Celestial Mechanics and Dynamical Astronomy \textbf{84} (2002) (3), 223-229	


\bibitem[StFo03]{StFo03}C. Stoica and A. Font, Global dynamics in the singular logarithmic potential, Journal of Physics A: Mathematical and General,  36 (2003) 7693-7714

\bibitem[Su07]{Su07} K. Sundman Recherches su le probl$\grave{\text{e}}$me des trois corps, Acta Societatis Scientierum Fennicae \textbf{34} (1907)

  \bibitem[VWD12]{VWD12} S. R. Valluri, P. A. Wiegert, J. Drozd and M. Da Silva,   A study of the orbits of the logarithmic potential for galaxies,  Mon. Not. R. Astron. Soc. \textbf{427} (2012) 2392–2400  

\bibitem[vdM21]{vdM21} J. C. van der Meer  Reduction and regularization of the Kepler problem
Celestial Mechanics and Dynamical Astronomy \textbf{133} (2021)

\bibitem[WilYor73]{WilYor73}
F.W.Wilson and R. Easton, Lyapunov functions and isolating blocks, J. Diff. Eq. 13 (1973), 106-123

\bibitem[LZ15]{LZ15} L Zhao, Kustaanheimo–Stiefel Regularization and the Quadrupolar Conjugacy, 
Regul. Chaotic Dyn., \textbf{20} (2015), 19-36   



\end{thebibliography}
\end{document}